\documentclass[aps,amssymb,floatfix,prd,amsmath,preprintnumbers]{revtex4}
\usepackage{xcolor}
\usepackage{graphicx}  % Include figure files
\usepackage{float}
\usepackage{dcolumn}   % Align table columns on decimal point
\usepackage{bm}% bold math
\begin{document}
%\input epsf.tex
%%%%%%%%%%%%
%%%%%%%%%%%

\title{\bf  Investigating the physical and geometrical parameters of the cosmological models with anisotropic background}

\author{B. Mishra \footnote{Department of Mathematics, Birla Institute of Technology and Science-Pilani, Hyderabad Campus, Hyderabad-500078, India, E-mail:bivudutta@yahoo.com },  S. K. Tripathy\footnote{Department of Physics, Indira Gandhi Institute of Technology, Sarang, Dhenkanal, Odisha-759146, India, E-mail:tripathy\_ sunil@rediffmail.com}
}\affiliation{ }

\begin{abstract}
In this paper, we have investigated some accelerating cosmological models at the backdrop of an anisotropic metric in an extended gravity theory. Two viable cosmological models one with a little rip behaviour and the other with a hyperbolic form of Hubble parameter have been constructed.  The dynamical aspects of the models along with some physical and geometrical parameters are analysed. Both the models presented here evolve in the phantom-like region and overlap with $\Lambda$CDM model at late times. We carried out a geometrical diagnosis of the model to show the viability of the models. 
\end{abstract}
\maketitle
\textbf{PACS number}: 04.50kd.\\
\textbf{Keywords}:  Extended gravity, Anisotropic universe, Little Rip, Hyperbolic function, Energy conditions.
\section{Introduction} 
Friedmann equations can be obtained when the Einstein's field equations are imposed on the Robertson-Walker space-time, however this path leads us directly to the meeting of one of the greatest questions in today's research in cosmology. At the end of twentieth century, we were presented with the fact that the Universe was expanding at an accelerated rate \cite{Riess98,Perlmutter99}, which implies that the stress energy component of Friedmann's equations must have negative pressure to explain such behavior. This exotic matter content is interpreted in literature in different ways (e.g., vacuum energy, scalar fields, Chaplygin gas, among others) depending on the model to which the analysis is made. The standard theoretical model supports this issue by imposing a ``subtle'' amount about $ 68.3 \% $ of the cosmic constituent on the form of a cosmological constant $\Lambda$. The so-called dark energy carries within it a repulsive behaviour to drive the Universe in an accelerated manner. Several efforts were made on the perspective of solidification of the model amidst a natural aspirant coming from particle physics, but so far without success, keeping the nature and cosmological origin of dark energy a puzzle which promotes strong debates in the academy. This brought some discomfort to the scientific community, which led to the search of alternative solutions to the issue. The prime motive is to modify General Relativity (GR) that demonstrates inefficiency at least at large scales. In these modified gravity theories, explanations to the late time cosmic acceleration  emerge from modified gravity dynamics rather than the dark energy.\\

In recent times, there have been a lot of modified gravity theories proposed in literature \cite{Capozziello03,Nojiri03, Carrol04, Satiriou10}. In these new gravity models, the geometrical action is modified by replacing the action of GR by a different one without incorporating any exotic matter fields. In 2011, Harko et al.\cite{Harko11} proposed $f(R,T)$ theory by considering a coupling of matter component and geometry and thereby geometrically modified the gravitational action. The matter geometry coupling within the gravitational action plays a significant role in providing a theoretical explanation to the late time cosmic acceleration \cite{Tripathy19, Tripathy20}. Although, this theory suffers from some observational issues such as growth of inhomogeneous structure \cite{alva13} and an issue like the non conservation of energy-momentum \cite{Shabani2017}. This extended gravity theory has gained a lot of research attention in recent years because of its simple structure and ability to explain many issues in cosmology and astrophysics. Reddy et al. \cite{Reddy12} have studied the spatially homogeneous Bianchi type III space-time whereas Adhav \cite{Adhav12} have investigated Bianchi I space-time in $f(R,T)$ gravity.  Samanta \cite{Samanta13} derived the exact solutions of the field equations with Kantowski-Sachs space-time filled with perfect fluid in the framework of $f(R,T)$ theory of gravity. Moraes \cite{Moraes14} unified Kaluza-Klein extra-dimensional model with $f(R,T)$ gravity and obtained the results from the induced matter model application. Reddy et al. \cite{Reddy14} constructed and analyzed the cosmological model with Kantowski-Sachs space-time in presence of bulk viscous fluid and one-dimensional cosmic strings in $f(R, T)$ gravity. Shamir \cite{Shamir15} solved the modified field equations using a relationship between scalar expansion and shear scalar. Mishra et al. \cite{Mishra16} have studied the dynamical behaviour of the model with a presumed power law scale factor. In this geometrically modified gravity theory, Mishra and Vadrevu \cite{Mishra17} have constructed the cosmological models in a cylindrically symmetric space-time.  Velten and Carames \cite{Velten17} have shown the difficulties in explaining a viable and realistic cosmology in modified gravity, in fact challenged the viability of modified gravity.  Jimenez and Koivisto \cite{Jimenez17} have discussed the cosmological applications and vector distortion of modified theory of gravity. Mishra et al.\cite{Mishra18a} have assumed the matter as viscous fluid to study the anisotropic universe.  A general formalism has been developed with the hybrid scale factor to study the dynamical behavior of the anisotropic space-time in extended gravity \cite{Mishra18b}. \\

Olmo \cite{Olmo11} reviewed the modified theories of gravity in the Palatini approach and has shown the importance of going beyond the $f(R)$ models to know the phenomenological aspects related to dark energy and quantum gravity.  In the case of the metric $f(R,T)$ gravity, after taking the divergence of the gravitational field equations, it has been obtained that the energy-momentum tensor of the matter is not conserved. Similar to this case, the motion of the particles is not geodesic, and because of this  matter geometry coupling, an extra force arises. This force has the same expression as in the metric case, so no new physics is expected to arise during the motion of massive test particles in the Palatini formulation of $f(R,T)$ gravity \cite{Wu18}. Following this Bamba and Odintsov \cite{Bamba15} have reviewed inflationary cosmology in modified gravity and also explained the bounce cosmology in $f(R)$ gravity. Tripathy and Mishra \cite{Tripathy16} have discussed the vacuum solution in an anisotropic universe. They have also shown the non-occurrence of big rip singularity in a phantom model in extended gravity both in isotropic and anistropic universe \cite{Tripathy20}. Paliathanasis \cite{Paliathanasis17} has considered quantum corrections in the Starobinsky model of inflation with the action integral correspond to the $f(R)$ theory of gravity and derived the analytic solution. Mishra et al. \cite{Mishra18c} have indicated that the increase in cosmic anisotropy substantially affect the energy conditions whereas in \cite{Mishra18d}, the affect on cosmic dynamics has been shown. With a hyperbolic form for Hubble parameter, Esmaeili and Mishra \cite{Esmaeili18a} and Esmaeili \cite{Esmaeili18b} have shown that irrespective of the scaling constant, the matter content in the gravitational theory remains unchanged in $f(R,T)$ gravity.  Tarai and Mishra \cite{Tarai18} have incorporated the magnetic field in the matter content and observed that with the inclusion of magnetic field in $f(R,T)$ gravity, there is a substantial effect on the dynamical behaviour of the model.  Aygun et al. \cite{Aygun19} have investigated the cosmological model with different quadratic equation of state parameter.\\

The $f(R,T)$ theory has been widely applied to understand different issues in astrophysics \cite{Yousaf16, Das17, Deb18, Biswas20}. Carvalho et al. \cite{Carvalho17} have investigated the equilibrium configurations of white dwarfs in $f(R,T)$ gravity. Islam and Basu \cite{Islam18} have presented the interior solutions of distributions of magnetized fluid inside a sphere, embedded with exterior Reissner-Nordstrom metric in $f(R,T)$ gravity.  Abbas and Ahmed \cite{Abbas19} have investigated charged perfect fluid spherically symmetric gravitational collapse in $f(R,T)$ gravity, where the interior boundary of a star filled with the charged perfect fluid. In a thermodynamical approach, Sharif and Zubair \cite{Sharif12} have discussed the $f(R,T)$ gravity for FRW universe whereas with the same background Sharif et al. \cite{Sharif13} obtained the bound on the physical parameters to satisfy the energy conditions. Tretyakov \cite{Tretyakov18} introduced higher derivatives matter fields to further modify the  $f(R, T)$ gravity and discussed the stability conditions.  Ordines and Carlson \cite{Ordines19} have shown the changes in Earth’s atmospheric models that comes from $f(R,T)$ modified theory of gravity, in fact they suggested limits on $f(R,T)$ gravity from Earth’s atmosphere. \\

When the cosmic energy density remains constant or strictly increasing in future, based on the time asymptotics of the Hubble parameter $H(t)$, we can divide the possible fates of the universe into four categories. This can be categorized as cosmological constant,big rip, little rip (LR) and pseudo-rip (PR) respectively when $H(t)$=constant, $H(t)\rightarrow \infty$ at finite time, $H(t) \rightarrow \infty$ as time goes to infinity, and $H(t)\rightarrow$ constant as time goes to infinity \cite{Frampton12a}. Usually inconsistencies occur due to finite time future singularity.  In order to avoid such inconsistencies, there have been some proposals to delay the singularity \cite{Nojiri2011, Framp2012}.  In some models, the  DE density increases with time, the equation of state parameter evolves asymptotically from $\omega<-1$ to $-1$ rapidly so that there is effectively no finite time future singularity \cite{Framp2011, Framp2012, Asta2012}. The Little Rip (LR) model belongs to these category of models. Bamba et al. \cite{Bamba12} have demonstrated that the disintegration of bound structures for LR and PR cosmologies occurs in the same way as in gravity with corresponding dark energy fluid. Balakin and Bochkarev \cite{Balakin13} have focused in constraining dark energy relaxation parameter specifically the dark energy equation of state parameter to avoid big rip singularity. Moreover, they have shown that the Archimedean type coupling protects the Universe to get into big rip scenario. Makarenko et al. \cite{Makarenko13} have constructed cosmological models in the framework of Gauss-Bonnet modified gravity and tested the singularity with LR. Frampton and Ludwick \cite{Frampton13} have studied the cyclic cosmology from LR. G\`omez \cite{Gomez13} has shown the non-occurrence of LR and big rip  scenario in $f(R)$ gravity as the expansion force is too small to produce any significant effect on the local system.  Boko et al. \cite{Boko17} have studied viscous cosmology with big rip and little rip in $f(R)$ gravity. Lopez et al. \cite{Lopez17} discussed the quantization of big rip in the framework of GR and modified gravity in FLRW metric.  In the situation when the dark energy content is described by the phantom like fluid or phantom scalar field, Albarran et al. \cite{Albarran18} addressed the quantization of the model that induces the little sibling of the big rip abrupt event. Tripathy \cite{Tripathy14} has studied LR cosmologies in the backdrop of an anisotropic Universe. In a recent work, we have investigated some phantom LR models in the framework of $f(R,T)$ gravity \cite{Tripathy20}. Obukhov et al. \cite{Obukhov2013} have investigated LR cosmologies via inhomogeneous cosmic fluids. Brevik et al. \cite{Brevik12} have investigated some LR viscous models. Wei et al. \cite{Wei12} have studied the little rip, pseudo rip cosmology to understand the fate of the Universe and have shown that quasi-rip is having a unique feature being different from big rip, little rip and pseudo rip. Parnovsky \cite{Parnovsky15} studied the possible types of future singularities such as Big Squeeze and Little Freeze in isotropic and homogeneous models. Recently, Brevik and Timoshkin \cite{Brevik20}  have investigated some viscous LR models in the framework of brane cosmology.\\

In the present work, we have investigated an LR model along with a cosmological model constructed through a hyperbolic Hubble rate in the framework of $f(R,T)$ gravity theory. We have considered an anisotropic Bianchi $VI_h$ ($BVI_h$) metric for our investigation. Eventhough cosmic homogeneity and isotropy are mostly observed in the present Universe at large scale (of the order of $100h^{-1}~Mpc$), deviation from isotropy can not be ruled out. Cosmic structures such as voids and super clusters in the local Universe are observed (\cite{Tripathy14},\cite{Ghodsi16}  and references therein). The paper is organized as follow: in Sec II, we discuss the basic formalism of the $f(R,T)$ gravity theory and derive the dynamical parameters for an anisotropic $BVI_h$ Universe in terms of the Hubble parameter. In Sec III, little rip scale factor and hyperbolic form for Hubble parameter are incorporated to obtain the equation of state parameter and to understand its evolutionary aspect. Also in this section we have discussed some physical properties of the models. In Sec IV, we have presented the energy conditions for both the constructed cosmological  models. The geometrical diagnostic analysis are carried out in section V. The conclusions of the research work have been given in Sec VI. In the present work, we use the natural system of units with $G=c=1$. where $G$ is the Newtonian gravitational constant and $c$ is the speed of light in vacuum.

\section{Basic Formalism and Dynamical Parameters}

\vspace{0.5cm}
Harko et al. \cite{Harko11} proposed the Einstein-Hilbert action for $f(R,T)$ gravity theory in the form 
\begin{equation} \label{eq:1}
S=\int d^4x\sqrt{-g}\frac{1}{16\pi} f(R,T)+\int d^4x\sqrt{-g}  \mathcal{L}_m .
\end{equation}
The action of $f(R,T)$ gravity as in \eqref{eq:1} is different from the action of GR. The Ricci scalar $R$ in the action of GR has been replaced with the function $f(R,T)$ in the action \eqref{eq:1}. $T$ is the trace of the energy momentum tensor $T_{\mu\nu}$. This coupling of matter and geometry leads to the non-vanishing divergence of $T_{\mu\nu}$. Moreover, due to this matter and geometry coupling, we may encounter with a strong reason for the cosmic acceleration issue. Harko et al. \cite{Harko11} proposed three different functional approaches to $f(R,T)$ such as (i) $f(R,T)=R+2f(T)$(ii) $f(R,T)=f_1(R)+f_2(T)$ and (iii)$f(R,T)=f_1(R)+f_2(R)f_3(T)$. It can be observed that the first approach $f(R,T)=R+2f(T)$ can be reduced to GR with some conditions. Here we have restricted ourselves to the first choice with the functional $f(R,T)=R+2\gamma T$, where $\gamma$ is a coupling constant. The value of $\gamma$ may be decided from certain physical basis. One should note that for $\gamma=0$, the features of GR can be obtained.\\

The matter Lagrangian $\mathcal{L}_m$ can be chosen in many different ways. It is worthy to mention here that the exact form of $\mathcal{L}_m$ is one of the fascinating theoretical problem in GR. One possible choice of the matter Lagrangian is $\mathcal{L}_m=-p$, $p$ being the pressure of the cosmic fluid,  to  derive the equation of motion of test fluids in standard GR \cite{Schutz70,Brown93}. The extra force, which is one of the distinguishing features of modified gravity theories with geometry-matter coupling, identically vanishes \cite{Bertolami08,Harko14,Sotiriou08}. It is notable to mention here that, the usual continuity equation is not satisfied for the $f(R,T)$ field equations. So the covariant derivative of the energy momentum tensor is not null in general. Alvarenga et al. \cite{alva13} have suggested a function that represents the unique Lagrangian which satisfies the continuity equation. Gomez et al. \cite{Gomez16} have reviewed the teleparallel extension of GR and presented the $f(T,\mathcal{T})$ gravity theory by simultaneously imposing the standard energy momentum conservation equation. This provides a theoretical prior on the specific forms of the Lagrangian that is analogous to \cite{alva13}. In view of the above, here we have considered the form of the matter Lagrangian as $\mathcal{L}_m=-p$. Consequently, the field equation of $f(R,T)$ gravity can be derived as,

\begin{equation} \label{eq:2}
f_R (R) R_{\mu\nu}-\frac{1}{2}f(R)g_{\mu\nu}= \left[8\pi +f_T(T)\right]T_{\mu\nu}+\left[f_T(T)p+\frac{1}{2}f(T)\right]g_{\mu\nu}+\left(\nabla_{\mu} \nabla_{\nu}-g_{\mu\nu}\Box\right)f_R(R).
\end{equation}

Here, $f_R=\frac{\partial f(R)}{\partial R} =1$ and $f_T=\frac{\partial f(T)}{\partial T}=2\gamma$. So the field eqns. \eqref{eq:2} reduce to 
\begin{equation} \label{eq:3}
R_{\mu\nu}-\frac{1}{2}Rg_{\mu\nu}= \kappa T_{\mu\nu} + \Lambda_{eff}(T)~ g_{\mu\nu},
\end{equation}
where $\kappa=2\gamma+8\pi$ and $\Lambda_{eff}(T)=2\gamma\left(p+\frac{1}{2}T\right)$. One should note that, the above equation \eqref{eq:3} looks like the GR field equation with a cosmological constant term $\Lambda_{eff}(T)$. The difference is that, this effective cosmological constant is not a constant rather varies dynamically depending on the behaviour of the matter field. We have considered here the energy momentum tensor as

\begin{equation}\label{eq:4}
T_{\mu\nu}=(\rho+p)u_{\mu}u_{\nu} - pg_{\mu\nu}-\xi x_{\mu}x_{\nu},
\end{equation}
with $u^{\mu}u_{\mu}=-x^{\mu}x_{\mu}=1, u^{\mu}x_{\mu}=0$. where $\rho=\rho_p+\xi$ is the energy density; $\rho_p$ and $\xi$ respectively represent particle energy density and string energy density. Since we are interested to study the cosmological model with additional anisotropic source, therefore we have considered  one dimensional cosmic string which has the contribution to the anisotropic nature of the cosmic fluid.

In GR, usually the Friedman models ensures the conservation of the energy-momentum tensor $\nabla^{\mu}T_{\mu\nu}=0$. But in modified gravity theories, we encounter a different situation. Taking a covariant derivative of eqn. \eqref{eq:2}, it can be obtained that \cite{Harko14, alva13, bar14, Das17} 
\begin{equation}
\nabla^{\mu}T_{\mu\nu}=\frac{f_T(T)}{8\pi-f_T(T)}[(T_{\mu\nu}+\Theta_{\mu\nu})\nabla^{\mu} ln f_T(T)+ \nabla^{\mu}\Theta_{\mu\nu} -\frac{1}{2}g_{\mu\nu}\nabla^{\mu}T], \label{eq:3a}
\end{equation}
where $\Theta_{\mu\nu} =g^{\alpha \beta} \frac{\delta T_{\alpha \beta}}{\delta g^{\mu\nu}}$ which for the specific choice of the matter Lagrangian $\mathcal{L}_m=-p$ becomes $\Theta_{\mu\nu}=-pg_{\mu\nu}-2T_{\mu\nu}$.
With the substitution of $f(R,T)=R+2\gamma T$, eqn. \eqref{eq:3a} reduces to
\begin{equation}
\nabla^{\mu}T_{\mu\nu}=-\frac{2\gamma}{8\pi+2\gamma} \left[\nabla^{\mu}(pg_{\mu\nu})+\frac{1}{2}g_{\mu\nu} \nabla^{\mu}T\right].\label{eq:3b}
\end{equation}

It is obvious that, for $\gamma=0$, we have $\nabla^{\mu}T_{\mu\nu}=0$. But for a non-vanishing value of $\gamma$, there is violation of energy-momentum conservation ($\nabla^{\mu}T_{\mu\nu}\neq 0$). In modified gravity theories, the non-conservation of the energy-momentum arises due to non-unitary modifications of quantum mechanics \cite{Josset2017}. In fact, Josset and Perez have shown that a non-conservation of energy-momentum leads to an effective cosmological constant which may be responsible for cosmic acceleration \cite{Josset2017}.

In order to investigate the dynamical behaviour of the Universe, we consider Bianchi $VI_h$ ($BVI_h$ ) space-time in the form

\begin{equation}\label{eq:5}
ds^2 = dt^2 - A_1^2dx^2- A_2^2e^{2x}dy^2 - A_3^2e^{2h x}dz^2,
\end{equation}
where $A_i=A_i(t)$, for $i=1,2,3$ and $\alpha$ is constant. Following Ref. \cite{Tripathy15}, we have assumed here $h=-1$. Now the field eqns. with the $BVI_{-1}$ space-time \eqref{eq:5} can be obtained as  
\begin{equation} \label{eq:6}
\frac{\ddot{A_2}}{A_2}+\frac{\ddot{A_3}}{A_3}+\frac{\dot{A_2}\dot{A_3}}{A_2A_3}+ \frac{1}{A_1^2}= -\beta p+\beta\xi +\gamma\rho,   
\end{equation}
\begin{equation} \label{eq:7}
\frac{\ddot{A_1}}{A_1}+\frac{\ddot{A_3}}{A_3}+\frac{\dot{A_1}\dot{A_3}}{A_1A_3}- \frac{1}{A_1^2}=-\beta p + \gamma\xi+\gamma\rho,   
\end{equation}
\begin{equation} \label{eq:8}
\frac{\ddot{A_1}}{A_1}+\frac{\ddot{A_2}}{A_2}+\frac{\dot{A_1}\dot{A_2}}{A_1A_2}- \frac{1}{A_1^2}=-\beta p +\gamma\xi+\gamma\rho,  
\end{equation}
\begin{equation} \label{eq:9}
\frac{\dot{A_1}\dot{A_2}}{A_1A_2}+\frac{\dot{A_2}\dot{A_3}}{A_2A_3}+\frac{\dot{A_3}\dot{A_1}}{A_3A_1}-\frac{1}{A_1^2}= -\gamma p+\gamma \xi+ \beta \rho,   
\end{equation}
\begin{equation} \label{eq:10}
\frac{\dot{A_2}}{A_2}=\frac{\dot{A_3}}{A_3}.
\end{equation} 
 
Here $\beta=8\pi+3\gamma$ and an over dot on a field variable denotes the ordinary derivative with respect to time. Since we wish to study the dynamics of the Universe with an assumed scale factor, we express the set of field eqns. \eqref{eq:6}-\eqref{eq:10} with respect to directional Hubble rates. We can consider the directional Hubble rates as: $H_x=\frac{\dot{A}_1}{A_1},H_y=\frac{\dot{A}_2}{A_2},H_z=\frac{\dot{A}_3}{A_3}$. Subsequently the mean parameter with respect to the scale factor $a$ and directional Hubble rate can be expressed as $H=\frac{\dot{a}}{a}=\frac{1}{3}(H_x+H_y+H_z)=\frac{1}{3}(H_x+2H_y)$. Suitably absorbing the integrating constant, eqn.\eqref{eq:10} produces the relation $H_y=H_z$. In order to obtain a functional form for the dynamical parameters, in this research, we have assumed a proportional relation between the amplitude of shear scalar $\sigma$ and the Hubble rate, which consequently resulted in $H_x=kH_y$, where $k$ is a constant.  This further provides anisotropic relationship between the scale factors. Tajahmad \cite{Tajahmad18} has considered the similar relation to  reconstructed the $F(T)$ gravity. Most recent, Tajahmad \cite{Tajahmad20} has shown that late time accelerated expansion arisen from gauge fields in an anisotropic background. It can be noted that for $k$ becomes unity, the space-time reduces to be isotropic. Rodriques et al. \cite{Rodriques14} and Bitterncourt et al. \cite{Bitterncourt17} have presented a detail study of isotropization of Bianchi type metrics. Now, the set of field eqns. \eqref{eq:6}-\eqref{eq:10} can be expressed as 

\begin{equation}
G_{11}(H)=-\beta p+\beta\xi +\gamma\rho
\end{equation}
\begin{equation}
G_{22}(H)=-\beta p + \gamma\xi+\gamma\rho
\end{equation}
\begin{equation}
G_{33}(H)=-\beta p +\gamma\xi+\gamma\rho
\end{equation}
\begin{equation}
G_{44}(H)=-\gamma p+\gamma \xi+ \beta \rho
\end{equation}
where,
\begin{eqnarray}
G_{11}(H)&=& \left(\frac{6}{(k+2)^2}\right)\dot{H}+\left(\frac{27}{(k+2)^2}\right)H^2+a^{-\left(\frac{6k}{k+2}\right)},\\
G_{22}(H) &=& G_{33}(H)= 3\left(\frac{k+1}{k+2}\right)\dot{H}+ 9\left(\frac{k^2+k+1}{(k+2)^2}\right)H^2-a^{-\left(\frac{6k}{k+2}\right)},\\
G_{44}(H)&=& 9 \left(\frac{2k+1}{(k+2)^2}\right)H^2- a^{-\left(\frac{6k}{k+2}\right)}.
\end{eqnarray}

Since $G_{22}=G_{33}$, henceforth we will use $G_{22}$ for both. Now, by doing some algebraic manipulations, we can obtain the pressure, energy density and string tension density in Hubble parameter form as:
\begin{eqnarray}
p&=&\left(\frac{\gamma}{\beta^2-\gamma^2}\right)[G_{11}(H)-G_{22}(H)+G_{44}(H)]-\left(\frac{\beta}{\beta^2-\gamma^2}\right)[G_{22}(H)],\\
\rho&=& \left(\frac{\beta}{\beta^2-\gamma^2}\right)[G_{22}(H)]-\left(\frac{\gamma}{\beta^2-\gamma^2}\right)[G_{11}(H)],\\
\xi&=&\left(\frac{1}{\beta-\gamma}\right)[G_{11}(H)-G_{22}(H)].
\end{eqnarray}

Subsequently, we can derive the equation of state parameter (EoS)  $\omega=\frac{p}{\rho}$ as
\begin{equation}
\omega = \frac{\gamma[G_{11}(H)-G_{22}(H)+G_{44}(H)]-\beta[G_{22}(H)]}{\beta[G_{22}(H)]-\gamma[G_{11}(H)]}.
\end{equation}
The effective cosmological constant ($\Lambda_{eff}$) can be obtained as

\begin{equation}
\Lambda_{eff}=\left(\frac{\gamma}{\beta+\gamma}\right)[G_{11}(H)+G_{44}(H)].
\end{equation}

\section{Cosmological Models with Scale Factors}
In the previous section, we have derived the physical parameters in term of Hubble parameter to assess the dynamics of the Universe and the background cosmology of the model. This can be achieved through the introduction of scale factors. Several scale factors are being incorporated in some  recent works e.g. power law \cite{Berman83}, hybrid scale factor  \cite{Mishra15}, bouncing scale factor \cite{Brandenberger11,Bars12,Tripathy19}, hyperbolic scale factor \cite{Esmaeili18a,Esmaeili18b} etc. In this paper, we have used the LR scale factor and a scale factor generated from a hyperbolic form of Hubble parameter to understand the  background cosmology and the issue of cosmic speed up phenomena. 

\subsection{Case-I (Little Rip)}

An important feature of the new cosmology is the big rip phenomenon. It means a singularity of the Universe to be encountered in a finite time \cite{Caldwell02}. Mathematically, we can understand this as the divergent integrals that follows from the Friedmann equations. The future singularity phenomenon in a soft variant is called the little rip. This has been characterized by an energy density that increases with time, however in an asymptotic sense, an infinite time is required to achieve the singularity \cite{Frampton12a}. Since we are interested to find the singularity if any in this cosmological model, we have assumed the little rip scale factor. Here,  we have considered an LR model with the Hubble parameter as, $H=Ae^{\lambda t}$, where $A>0$ and $\lambda$ are constant parameters. One should note that, for positive values of $\lambda$, the LR behaviour is manifested by the Hubble parameter and for negative values of $\lambda$, we get a usual Hubble parameter that decreases with cosmic time. In the present work, mostly we are interested in a model with LR behaviour that describes a late time universe and therefore consider a positive value for $\lambda$.  The scale factor for such an LR behaviour can be expressed as $a=a_0~exp\left[\frac{A}{\lambda}\left(e^{\lambda t}-e^{\lambda t_0}\right)\right]$. Here, $a_0$ is the scale factor at the present epoch $t_0$ and is considered to be 1. The deceleration parameter for this LR model can be expressed as $q=-1+\frac{d}{dt}\left(\frac{1}{H}\right)=-1-\frac{\lambda}{A}e^{-\lambda t}$. One should note that, the Hubble rate increases exponentially with time and diverges as $t\rightarrow \infty$. The deceleration parameter remains negative throughout the cosmic evolution. For given values of the model parameters $A>0$ and $\lambda>0$, the deceleration parameter increases from some large negative values to $q=-1$ at late times of cosmic evolution. It is to note here that, since we have $e^{-\lambda t}>0$, the deceleration parameter remains negative for positive values of $\lambda$ and it means an ever-accelerating universe. If one wishes to have a decelerated epoch, then the negative values of $\lambda$ may be considered. Therefore, the aforementioned type of scale factor belongs to a decelerated or an accelerated era according to the choice of $\lambda$. In FIG. 1, we have shown the behaviour of the deceleration parameter for the LR model as a function of the cosmic time. In the figure, we have only shown the accelerated era. It is worth to mention here that, the motivation behind an LR model is to avoid singularity at finite time scale and therefore, in this model, we cannot demonstrate the transit behaviour of the Universe from a decelerated phase of expansion to an accelerated one. 

\begin{figure}[!h]
\includegraphics[width=90mm]{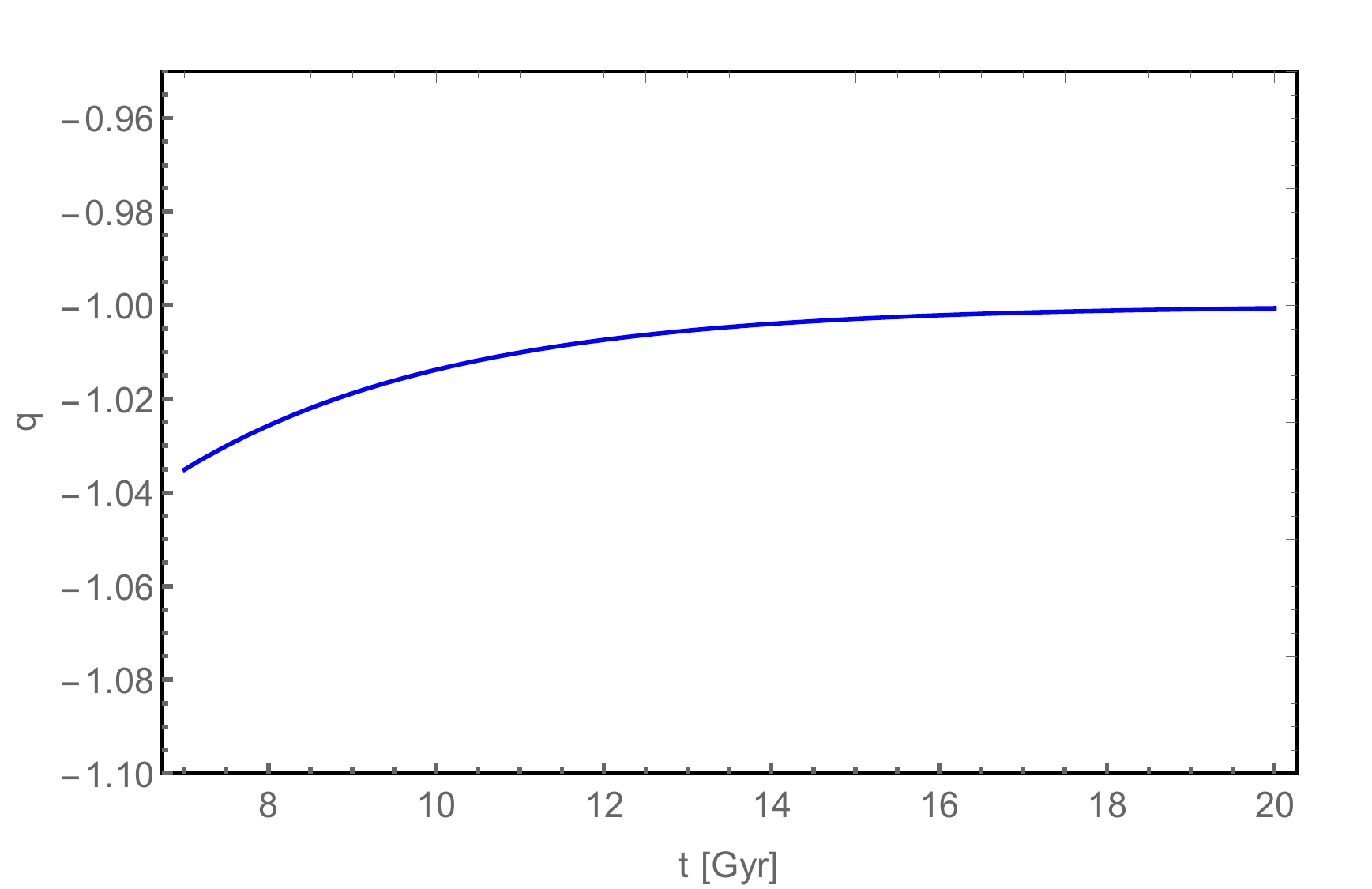}
\caption{Deceleration Parameter ($q$) for the LR model is plotted as a function of cosmic time. Here we have considered $\lambda=0.3122$ and $A=1$. The value of $q$ in the present epoch is $-1.004$.}
\end{figure}

For this LR model, the pressure, energy density, and string tension density can be obtained as  
\begin{eqnarray} 
p&=&\frac{\gamma}{(\beta^2-\gamma^2)(k+2)^2}\left[3(k^2+k-2)\lambda A e^{\lambda t}+9(k^2-k-3)(Ae^{\lambda t})^2\right]\nonumber\\
&-&\frac{\beta}{(\beta^2-\gamma^2)(k+2)^2}\left[3(k^2+3k+2)\lambda Ae^{\lambda t}+ 9(k^2+k+1)(Ae^{\lambda t})^2\right],\\ 
\rho&=&\frac{\beta}{(\beta^2-\gamma^2)(k+2)^2}\left[9(2k+1)(Ae^{\lambda t})^2\right]\nonumber\\
&-&\frac{\gamma}{(\beta^2-\gamma^2)(k+2)^2}\left[6(k+2)\lambda Ae^{\lambda t}+ 27(Ae^{\lambda t})^2 \right],\label{eq:24}\\ 
\xi=&-&\frac{3(k-1)}{(8\pi+2\gamma)(k+2)}\left[\lambda Ae^{\lambda t} +3(Ae^{\lambda t})^2\right].
\end{eqnarray}

It is obvious that, the evolutionary aspect of the pressure, energy density and string tension density depend on the model parameters $\gamma, \lambda, A$ and the anisotropy parameter. In the present work, we have  considered the anisotropy parameter to be $k=1.0000814$ which corresponds to an average anisotropy $\mathcal{A}=\frac{1}{3}\Sigma_{i=1}^{3}\left(1-\frac{H_i}{H}\right)^2=4.91\times 10^{-10}$. Similar observational bounds on the cosmic anisotropy have been obtained and suggested in the literature \cite{Saadeh16, Jaffe06}. The parameters $\lambda$ and $A$ are chosen in such a manner to provide a suitable description of the deceleration parameter and the Hubble rate. For different plots representing the behaviour of the equation of state parameter and the effective cosmological constants, we have considered (for brevity) $A=1~km~s^{-1}~Mpc^{-1}$. With this value of the model parameter $A$, we get the Hubble parameter $H\simeq 74.32$ for $\lambda t_0\simeq 4.308$. Considering the present time scale of the Universe as $t_0\simeq 13.8$ Gyr, we can have $\lambda\simeq 0.3122$(Gyr)$^{-1}$. The deceleration parameter in the present epoch is obtained as $q_0=-1.004$ which is close to the value $q_0=-1.08\pm 0.29$ as constrained in a recent analysis \cite{Camarena20}. It should be mentioned here that, from local distance ladder measurement, Reiss et al. \cite{Reiss18} have found the value of the Hubble parameter at the present epoch as  $H=74.3\pm 1.42~km~s^{-1}~Mpc^{-1}$ .  Since there are no observational constraints available on the parameter $\gamma$, we chose it to be a free parameter so that the energy density for the present model remains positive throughout the cosmic evolution. In view of this, we have chosen four different values of $\gamma$ namely $\gamma=-1.53, -0.03, 1.47$ and $2.97$. The string tension density $\xi$ becomes a dynamical quantity and its time varying nature also depends on the model parameter $\gamma$ besides $k,A$ and $\lambda$. For the given values of the model parameters, the sting tension density comes out to be a negative quantity and decreases with cosmic time. Since $\xi$ is negative quantity for the present model, with an increase in the value of $\gamma$, it increases.

The EoS parameter $\omega=\frac{p}{\rho}$ for the LR model can be derived as

\begin{eqnarray}
\omega = -1+(\beta+\gamma)\left[\frac{3(k^2+3k+2)\lambda A e^{\lambda t}+9(k^2-k)A^2e^{2\lambda t}}{\gamma\left[6(k+2)\lambda Ae^{\lambda t}+27A^2e^{2\lambda t}\right]-\beta\left[9(2k+1)A^2e^{2\lambda t}\right]}\right]. 
\end{eqnarray}
The evolutionary behaviour of the EoS parameter obviously depends on the model parameters. In general $\omega$ dynamically evolves  in the phantom region ($\omega<-1$) and asymptotically reaches to a value of $\omega=-1$ at late times. In view of this, it can be inferred that the model overlaps with the $\Lambda$CDM model at late times of cosmic evolution. In FIG.2, we have shown the EoS parameter as function of cosmic time for four different values of $\gamma$. As can be observed from the figure, the value of $\gamma$ affects the rate of evolution in the EoS parameter. Higher is the value of $\gamma$, lower is the rate of growth for $\omega$. Interestingly all the curves of $\omega$ merge to assume the asymptotic value of $-1$. At the present epoch, the present model predicts the EoS parameter to be $\omega(t_0)=-1.001, -1002, -1.003$ and $-1.004$ respectively corresponding to the values of the parameter $\gamma=-1.53, -0.03, 1.47$ and $2.97$.
%*********************************
%******************************
\begin{figure}[!h]
\includegraphics[width=90mm]{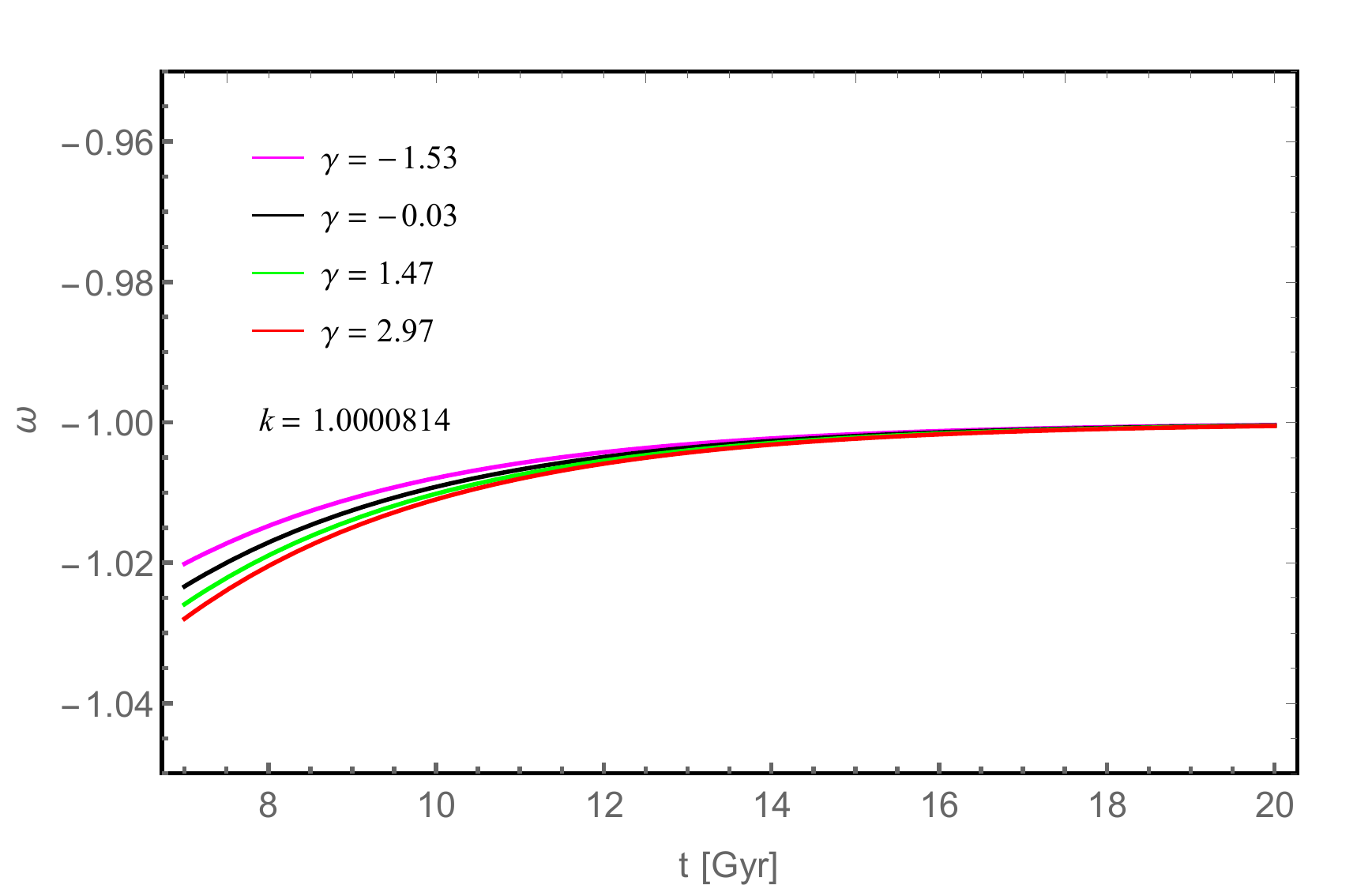}
\caption{The EoS Parameter ($\omega$) for the LR model is plotted as a function of cosmic time for four different values of $\gamma$.}
\end{figure}
%***************************************
%****************************************

The dynamical effective cosmological constant $\Lambda_{eff}$ for this model is obtained as
\begin{equation}
\Lambda_{eff}=\frac{\gamma}{(\beta+\gamma)(k+2)^2}\left[6(k+2)\lambda Ae^{\lambda t }+ 18(k+2)A^2e^{2\lambda t}\right].
\end{equation}
In FIG. 3, the effective cosmological constant is plotted for the four different values of $\gamma$ considered in the present work. Contrary to the behaviour of the usual cosmological constant with a time varying nature in dynamical dark energy models, in the present model, for negative values of $\gamma$, the effective cosmological constant becomes a negative quantity and decreases with the growth of cosmic time. For positive values of $\gamma$, it shows an increasing behaviour. It is worth to mention here that, the effective cosmological constant depends on the choice of the coupling constant $\gamma$. In case of $\gamma=0$, the $f(R,T)$ model reduces to GR and eventually, the effective cosmological constant vanishes. Also, it shoulders some burden of providing an accelerating model at late times.

%*********************************
%******************************
\begin{figure}[!h]
\includegraphics[width=90mm]{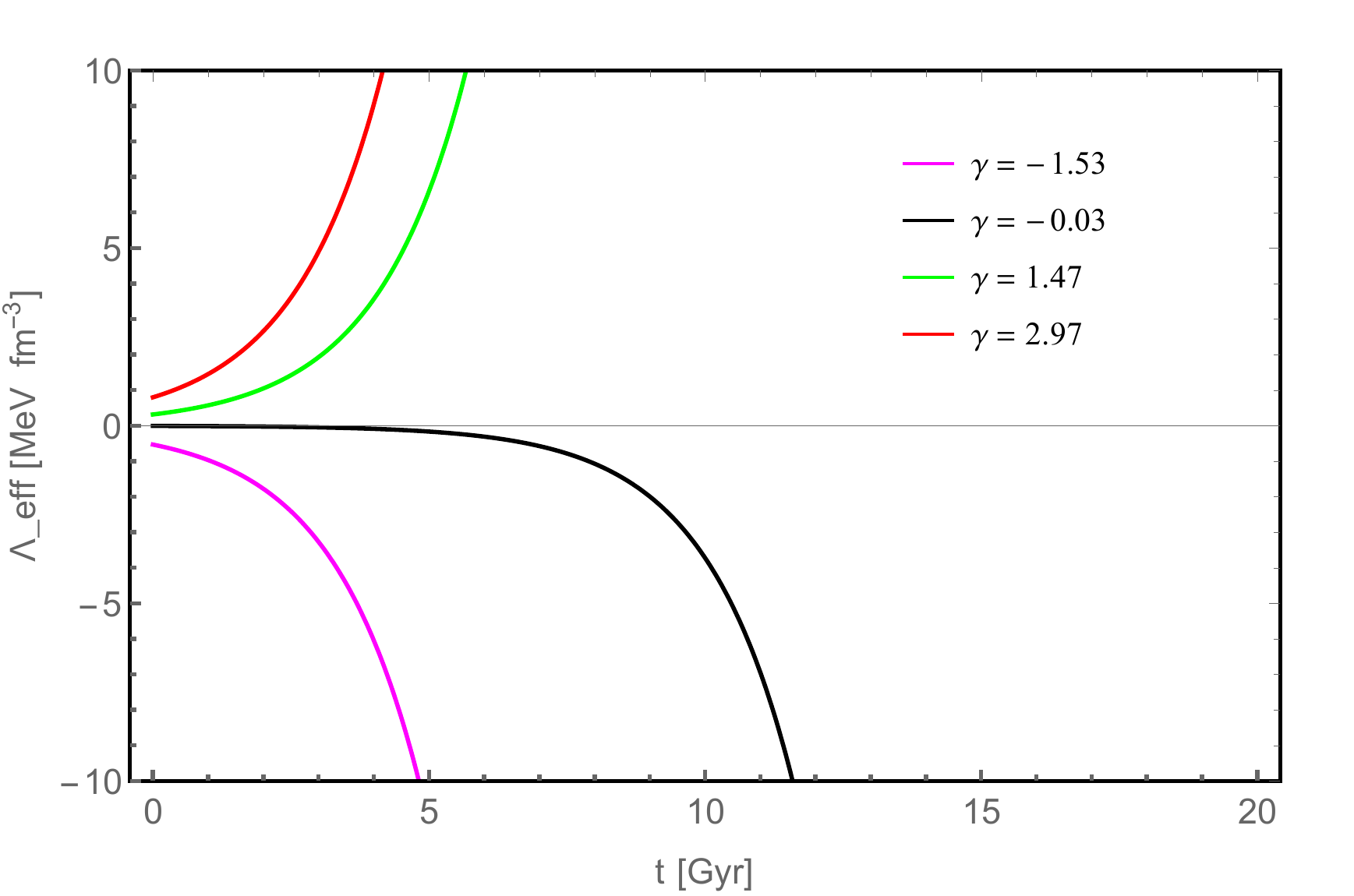}
\caption{The effective cosmological constant ($\Lambda_{eff}$) shown as function of cosmic time for four different values of the parameter $\gamma$ for the LR model.}
\end{figure}
\subsection{Case-II(Hyperbolic Form for Hubble Parameter)}

In this section, we wish to consider a hyperbolic form of Hubble parameter $H=\mathcal{H}_0\cosh{\mu t}$, where $\mathcal{H}_0$ and $\mu$ are the model parameters which need to be fixed from some physical basis. Hereafter, we designate this case as the hyperbolic model. Since $\cosh{\mu t}=\frac{e^{\mu t}+e^{-\mu t}}{2}$, this model has an extra term as compared to that of the LR model. The extra term in the Hubble parameter is expected to contribute to the cosmic dynamics. The model crosses the phantom divide at $t=0$. For $\mu t>>1$, $H\sim\frac{\mathcal{H}_0}{2}e^{\mu t}$ and the model behaves like the LR model. It can be noted that for non-phantom-like phase, for $t<0$, $\dot{H}<0$ and for phantom-like phase, for $t>0$,  $\dot{H}>0$. The scale factor for such a form of Hubble parameter becomes $a=a_0~exp\left[\frac{\mathcal{H}_0}{\mu}\sinh{\mu t}\right]$, $a_0$ being an integration constant. The deceleration parameter  for this model becomes $q=-1-\frac{\mu \sinh{\mu t}}{\mathcal{H}_0 \cosh^2 {\mu t}}$. As the hyperbolic function will always be positive, the sign of $q$ depends on $\mu$ and $\mathcal{H}_0$. Also, the inflation in the universe depends on the sign of the deceleration parameter. Nagpal et al. \cite{Nagpal19} have shown the cosmological implications of the model with the Hubble parameter as $H(t)=\frac{B\coth(B t)}{m}$ in a quadratic correction  geometric term of $f(R,T)$ gravity. Here $B$ and $m$ are constants. Biswas et al. \cite{Biswas12} have considered the Hubble parameter in the form of tangent hyperbolic function and studied the bounce and inflation in non-local higher derivative cosmology. 

The pressure, energy density, string energy density for the present hyperbolic model can be obtained as 

\begin{eqnarray} 
p&=&-\frac{\gamma}{(\beta^2-\gamma^2)(k+2)^2}\left[3(k^2+k-2)\mu \mathcal{H}_0 \sinh{\mu t}+9(k^2-k-3)(\mathcal{H}_0 \cosh{\mu t})^2\right]\nonumber\\ 
&-&\frac{\beta}{(\beta^2-\gamma^2)(k+2)^2}\left[3(k^2+3k+2)\mu \mathcal{H}_0 \sinh{\mu t}+9(k^2+k+1)(\mathcal{H}_0 \cosh {\mu t})^2\right],\\ 
\rho &=&\frac{\beta}{(\beta^2-\gamma^2)(k+2)^2}\left[9(2k+1)(\mathcal{H}_0 \cosh{\mu t})^2\right]\nonumber\\ 
&-&\frac{\gamma}{(\beta^2-\gamma^2)(k+2)^2}\left[6(k+2)\mu \mathcal{H}_0 \sinh {\mu t}+27(\mathcal{H}_0 \cosh{\mu t})^2\right],\\
\xi&=&-\frac{3(k-1)}{(8\pi+2\gamma)(k+2)}\left[\mu \mathcal{H}_0 \sinh {\mu t}+3(\mathcal{H}_0 \cosh {\mu t})^2\right].
\end{eqnarray}

As in the previous LR model, in this hyperbolic model, we have fixed up the model parameters from some physical basis. For brevity, we have considered, $\mathcal{H}_0=1~km~s^{-1}~Mpc^{-1}$. We obtain $H\simeq 74.3$ corresponding to $\mu t_0\simeq 5.00112$. Considering the present time scale of the Universe as $t_0\simeq 13.8$ Gyr, we can have $\mu\simeq 0.3624$ (Gyr)$^{-1}$.  In FIG. 4, we have shown $q$ as a function of cosmic time. The deceleration parameter initially decreases, attains a minimum and then rises asymptotically to become $-1$ at late times. At the present epoch its value is obtained to be $q_0=-1.005$. The minimum in the deceleration parameter occurs at $t\simeq 2.21$ Gyr. It is important to mention that, in the present work, we intend to model a late time accelerating Universe and use here a hyperbolic model which can not demonstrate the transition of the Universe from a decelerated phase of expansion to an accelerated one.  In view of this, we present the qualitative behaviour of the parameters with a care of their present values. Also, we have not investigated any bouncing behaviour at an initial epoch for the hyperbolic model. 

We consider the same value of the anisotropy parameter $k$  for the present hyperbolic model as that of the LR case. Also, we have chosen the values of $\gamma$ in such a manner that, the energy density in the present model remains positive throughout the cosmic evolution. The values of $\gamma$ that satisfy the above condition as considered in the present work are $-1.53, -0.03, 1.47$ and $2.97$. In this case also, for the given set of model parameters, the string tension density becomes a dynamically decreasing negative quantity and increases as the value of $\gamma$ increases.

\begin{figure}[!h]
\includegraphics[width=90mm]{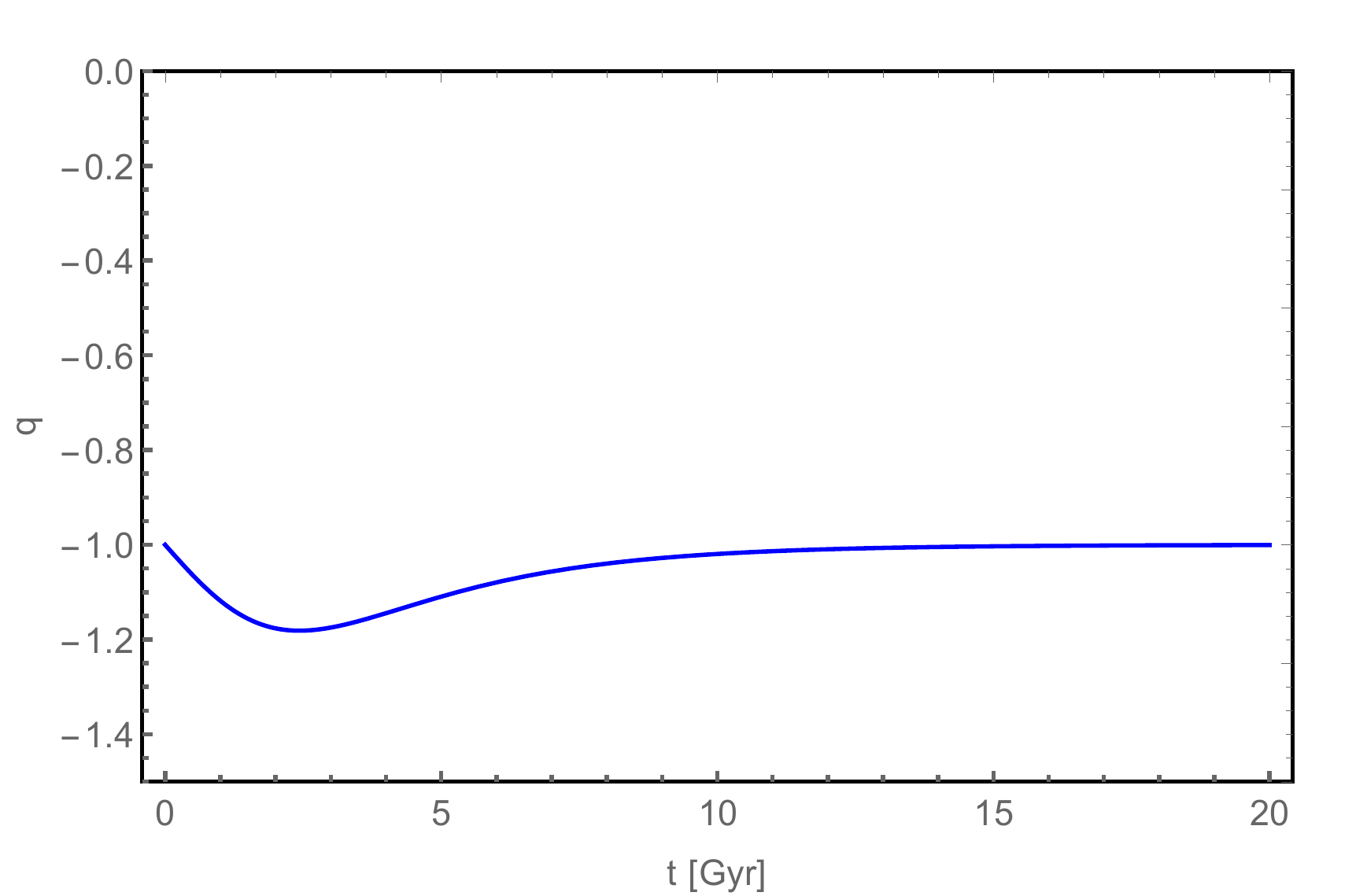}
\caption{Deceleration Parameter ($q$) for the hyperbolic model is plotted as a function of cosmic time. Here we have considered $\mu=0.3624$ and $\mathcal{H}_0=1$. The value of $q$ in the present epoch is $-1.004$.}
\end{figure}

For this hyperbolic model, we can obtain the EoS parameter and effective cosmological constant as
\begin{eqnarray}
\omega &=&-1 \\ 
&+& (\beta+\gamma)\left[\frac{3(k^2+3k+2)\mu \mathcal{H}_0 \sinh{\mu t}+9(k^2-k)(\mathcal{H}_0 \cosh{\mu t})^2}{\gamma\left[6(k+2)\mu \mathcal{H}_0 \sinh {\mu t}+27(\mathcal{H}_0 \cosh{\mu t})^2\right]-\beta\left[9(2k+1)(\mathcal{H}_0 \cosh {\mu t})^2\right]}\right],\\
\Lambda_{eff} &=&\frac{\gamma}{(\beta+\gamma)(k+2)^2}\left[6(k+2)\mu \mathcal{H}_0 \sinh {\mu t}+18(k+2)(\mathcal{H}_0 \cosh{\mu t})^2\right].
\end{eqnarray}
%%%%%%%%%%%%%%%%%%%%%%%
%*********************************
%******************************
\begin{figure}[!h]
\includegraphics[width=90mm]{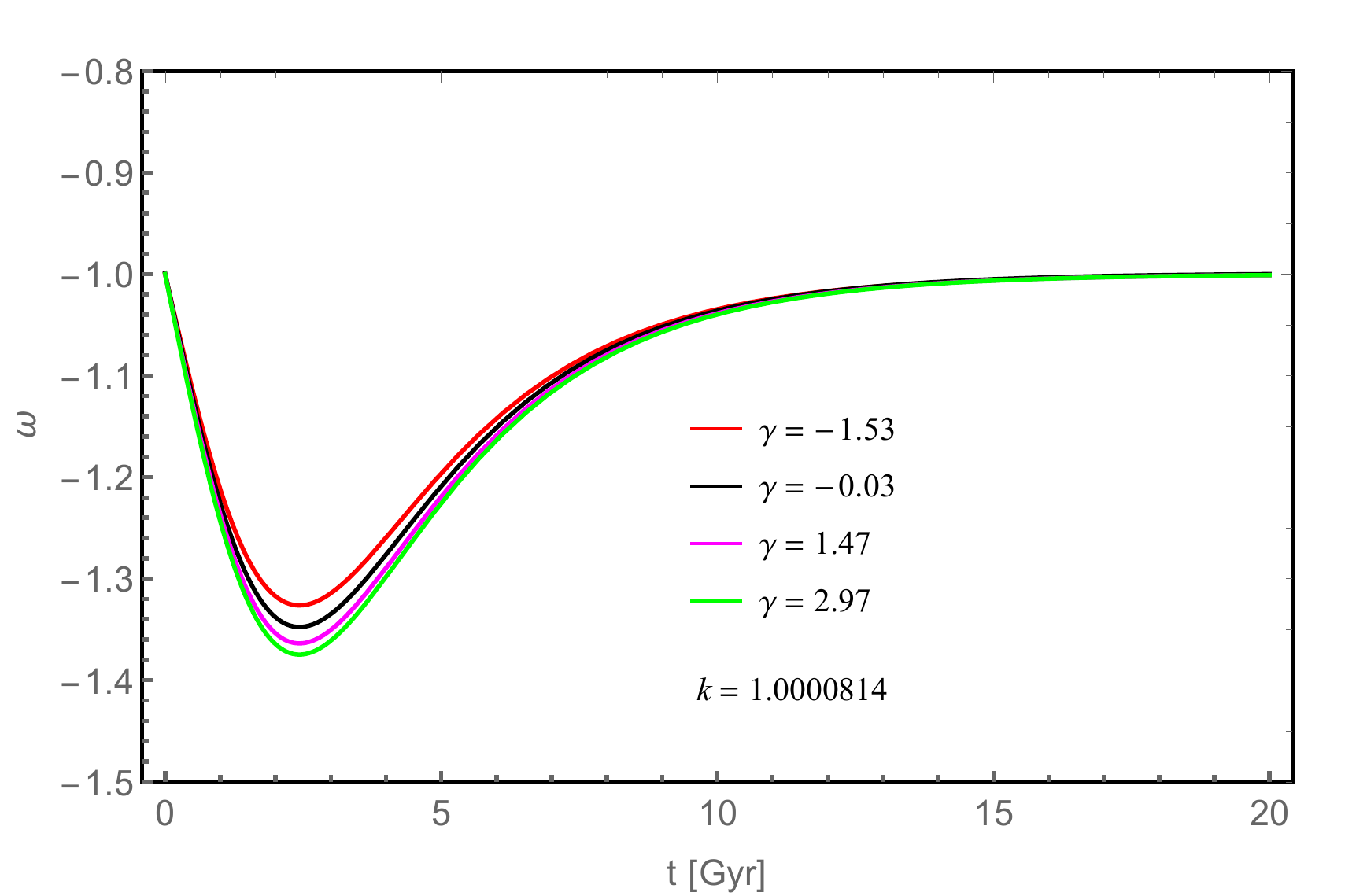}
\caption{The EoS Parameter ($\omega$) for the hyperbolic model is plotted as a function of cosmic time for four different values of $\gamma$.}
\end{figure}
%***************************************
%****************************************
%*********************************
%******************************
\begin{figure}[!h]
\includegraphics[width=90mm]{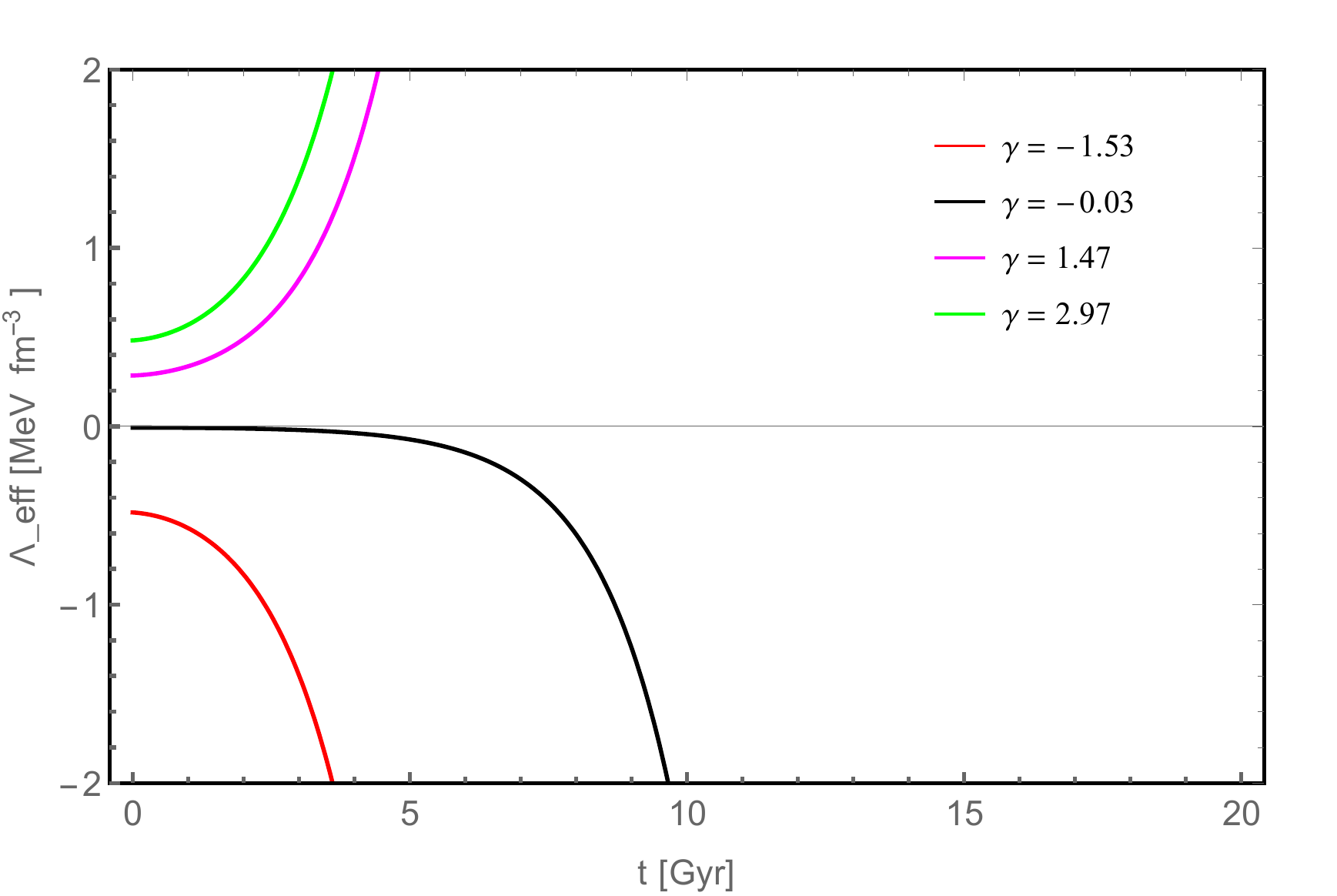}
\caption{The effective cosmological constant $\Lambda_{eff}$ for the hyperbolic model.}
\end{figure}
%***************************************
%****************************************

%
%
%
%\begin{figure}[!h]
%\minipage{0.50\textwidth}
%\includegraphics[width=65mm]{HYP_omegavst.pdf}
%\caption{The EoS Parameter ($\omega$) for the hyperbolic model.}
%\endminipage\hfill
%\minipage{0.50\textwidth}
%\includegraphics[width=65mm]{HYP_lambdavst.pdf}
%\caption{The effective cosmological constant $\Lambda_{eff}$ for the hyperbolic model.}
%\endminipage
%\end{figure}
%%%%%%%%%%%%%%%%%%%%%%%%

In FIG. 5, we have shown the EoS parameter for the hyperbolic function as a function of cosmic time for the representative values of the parameter $\gamma$. It is clear that, the EoS parameter is a negative quantity for the time zone considered in the present work. For all the values of $\gamma$,  it evolves down to a phantom-like phase initially and attains a minimum at around $t\simeq 2.21$ Gyr. Beyond this epoch, $\omega$ increases to reach the asymptotic value of $-1$. The value of $\gamma$ decides the depth of the $\omega$-well occurring in an initial  epoch. Higher is the value of $\gamma$, deeper is the well. However, for all the possible values of $\gamma$ considered in the work, the model overlaps with the $\Lambda$CDM model at times. The hyperbolic model predicts $\omega(t_0)=-1.008$ at the present epoch for all the values of $\gamma$.

In FIG. 6, the evolutionary aspect of the effective cosmological constant $\Lambda_{eff}$ for the hyperbolic model is shown. The general trend of this quantity is the same as that of the LR model. The effective cosmological constant becomes positive for positive choices of the $\gamma$ values and negative for negative $\gamma$. An increase in the magnitude of $\gamma$ increases the rate of dynamical evolution of $\Lambda_{eff}$.

\section{Energy Conditions}
Analysis of energy conditions is an integral part of the cosmological models as it add some additional constraints to the models \cite{Carroll04}.  The conditions are: (i) Null Energy Condition (NEC): $\rho+p\geq0$; (ii) Weak Energy Condition (WEC): $\rho+p\geq0$, $\rho\geq0$; (iii) Strong Energy Condition (SEC): $\rho+3p\geq0$; and (iv) Dominant Energy Condition (DEC): $\rho-p\geq0$. Sharif and Zubair \cite{Sharif13b} have suggested constraints to satisfy the power law solutions in order to aligned with the bounds prescribed by the energy conditions. Alvarenga et al. \cite{Alvarenga13b} have suitably adjusted the input parameters to satisfy the energy conditions. 

The energy conditions for the little rip (Case-I) scale factor can be derived as:

\begin{eqnarray}
\rho+p=&-&\frac{1}{(8\pi+2\gamma)(k+2)^2}\left[3(k^2+3k+2)\lambda Ae^{\lambda t}+9(k^2-k)(Ae^{\lambda t})^2\right],\\
\rho+3p=&-&\frac{\gamma}{(\beta^2-\gamma^2)(k+2)^2}\left[3(3k^2+5k-2)\lambda Ae^{\lambda t}+27(k^2-k-2)(Ae^{\lambda t})^2\right]\nonumber\\ 
&-&\frac{\beta}{(\beta^2-\gamma^2)(k+2)^2}\left[9(k^2+3k+2)\lambda Ae^{\lambda t}+ 9(3k^2+k+2)(Ae^{\lambda t})^2\right],\\ 
\rho-p=&&\frac{\gamma}{(\beta^2-\gamma^2)(k+2)^2}\left[3(k^2-k-6)\lambda Ae^{\lambda t}+9(k^2-k)(Ae^{\lambda t})^2\right]\nonumber\\ 
&+& \frac{\beta}{(\beta^2-\gamma^2)(k+2)^2}\left[3(k^2+3k+2)\lambda Ae^{\lambda t}+ 9(k^2+3k+2)(Ae^{\lambda t})^2\right].
\end{eqnarray}

Similarly, the energy conditions for the hyperbolic Hubble parameter (Case-II) can be obtained as,
\begin{eqnarray}
\rho+p=&-&\frac{1}{(8\pi+2\gamma)(k+2)^2}\left[3(k^2+3k+2)\mu \mathcal{H}_0 \sinh {\mu t}+9(k^2-k)(\mathcal{H}_0 \cosh {\mu t})^2\right],\\
\rho+3p=&-&\frac{\gamma}{(\beta^2-\gamma^2)(k+2)^2}\left[3(3k^2+5k-2)\mu \mathcal{H}_0 \sinh{\mu t}+27(k^2-k-2)(\mathcal{H}_0 \cosh{\mu t})^2\right]\nonumber\\ 
&-&\frac{\beta}{(\beta^2-\gamma^2)(k+2)^2}\left[9(k^2+3k+2)\mu \mathcal{H}_0 \sinh{\mu t}+ 9(3k^2+k+2)(\mathcal{H}_0 \cosh{\mu t})^2\right],\\ 
\rho-p=&+&\frac{\gamma}{(\beta^2-\gamma^2)(k+2)^2}\left[3(k^2-k-6)\mu \mathcal{H}_0 \sinh{\mu t}+9(k^2-k)(\mathcal{H}_0 \cosh{\mu t})^2\right]\nonumber\\ 
&+&\frac{\beta}{(\beta^2-\gamma^2)(k+2)^2}\left[3(k^2+3k+2)\mu \mathcal{H}_0 \sinh{\mu t}+ 9(k^2+3k+2)(\mathcal{H}_0 \cosh{\mu t})^2\right].
\end{eqnarray}

%%%%%%%%%%%%%%%%%%%%%%%
\begin{figure}[!h]
\minipage{0.50\textwidth}
\includegraphics[width=65mm]{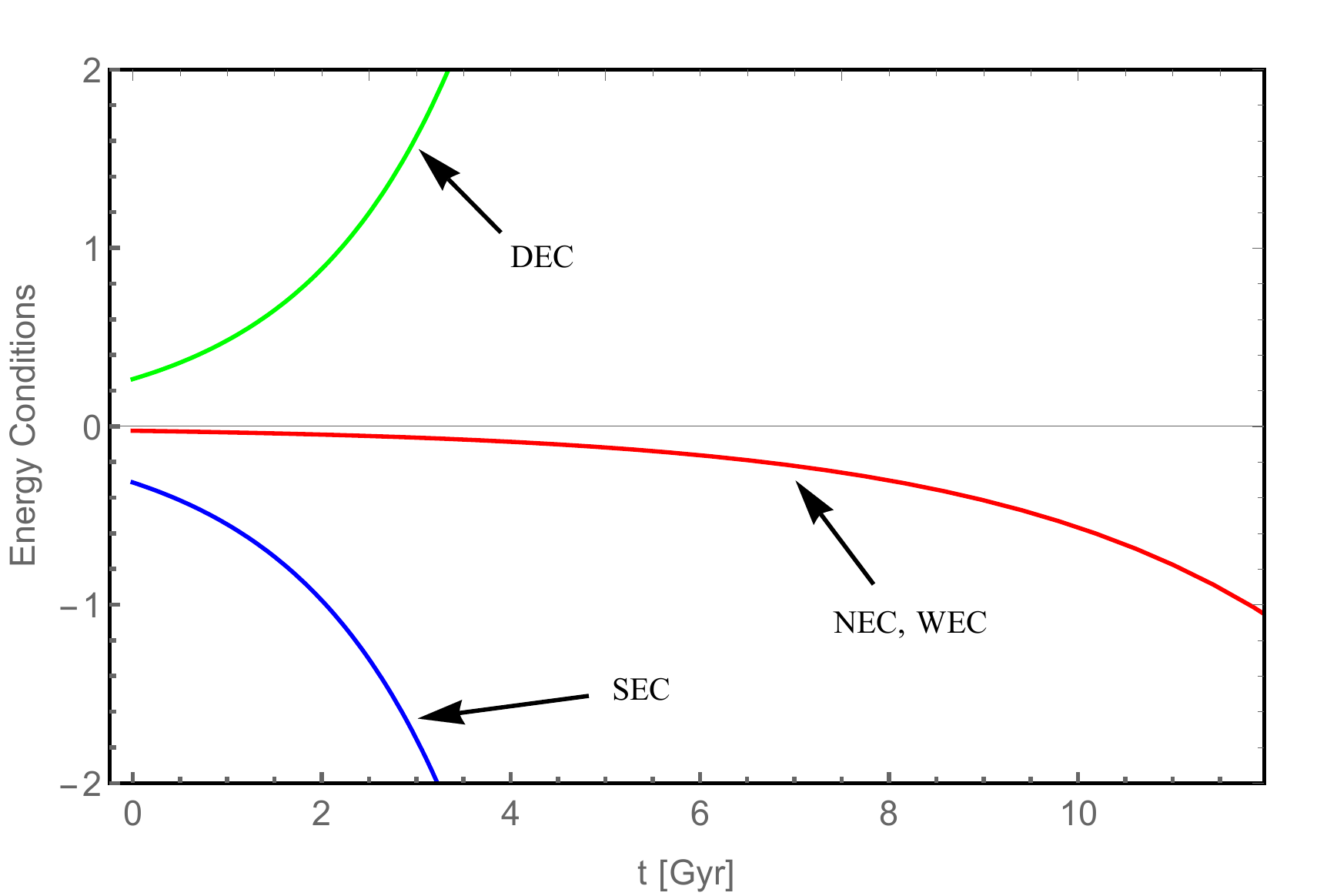}
\caption{Energy Conditions for the LR model.}
\endminipage\hfill
\minipage{0.50\textwidth}
\includegraphics[width=65mm]{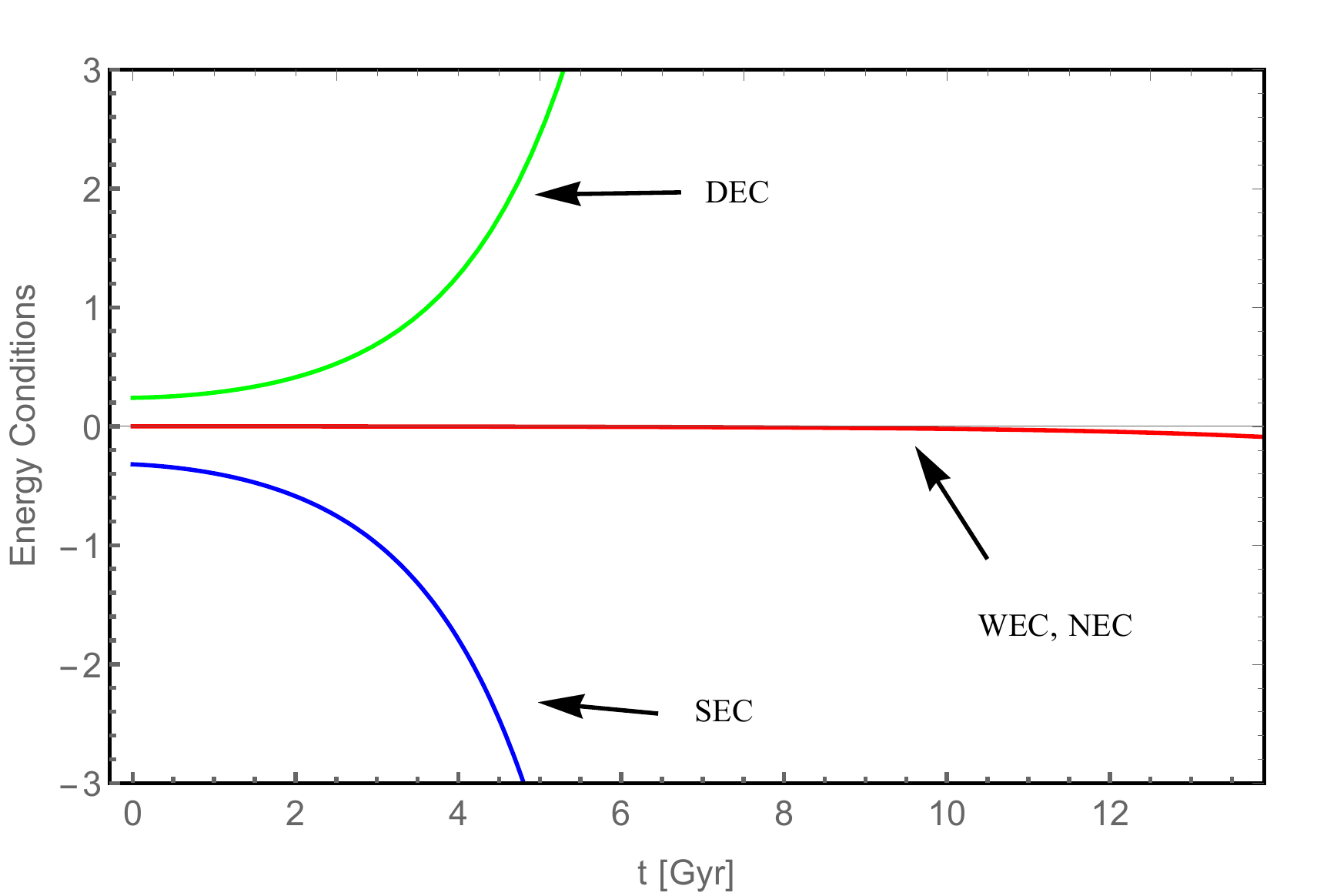}
\caption{Energy Conditions for the hyperbolic model.}
\endminipage\hfill
\end{figure}
%%%%%%%%%%%%%%%%%%%%%%%%

The behaviour of energy conditions are represented graphically for Case-I and Case II in FIG. 7 and FIG. 8 respectively. Since the value of $\gamma$ (within the limits of its consideration in the present work) affects only the stiffness of the EoS parameter without changing the general evolutionary behaviour, we have considered a representative value $\gamma=-0.03$ for plotting the figures of the energy conditions. The two models we have discussed, evolve in the phantom-like phase. Therefore it is expected that, except the DEC, all other energy conditions should be violated. The same features have been confirmed in the figures of the energy conditions for both the models. 

\section{Geometrical Analysis}

Here we will present the geometrical diagnostic of the models through the calculation of the state finder pair $\{r,s\}$ in the $r-s$ plane. Since there are a good number of models proposed in the literature with quite different behaviour of the Hubble and deceleration parameters, the state finder pair can provide a more sensitive and essential analysis to distinguish between dark energy models \cite{Sahni03}. Models such as $\Lambda$CDM, Quintessence (Q) and Kinessence (K), which have a parameter for the equation of state in the form, $\omega=-1$, $\omega<-1/3$ (condition for universe acceleration) and $-1<\omega<0$, respectively, but in (K) $\omega$ is not constant as in the previous cases, having a temporal dependence. We should also remember that the Q and K models are usually derived from a category of tracker fields \cite{Johri01}, minimally coupled dark energy (inflaton-like fields which converge along an evolutionary path). Besides  constant Quintessence models, there can also be dynamical Quintessence dark energy models where the equation of state parameter evolves with cosmic time.
It is worth noting the existence of transient models known as Quintom model, where its parameter $\omega$ can cross the phantom divide line  $(\omega =-1)$. These models are based on observational data as they restrict the $\omega$ parameter in the range $-1.6 <\omega<-0.8$ \cite{Bean02,Melchiiorri02}.\\

The state finder pair with respect to the Hubble parameter can be defined as  $r=\frac{\ddot{H}}{H^3}-(3q+2)$ and $ s= \frac{2(r-1)}{3(2q-1)}$. In the case of little rip model, the state finder pair $(r,s)$ can be obtained as 
\begin{eqnarray}
r &=& 1+\frac{\lambda^2+3\lambda Ae^{\lambda t}}{(A e^{\lambda t})^2},\\
s &=& -\frac{2}{3}\times\frac{(\lambda^2+3\lambda A e^{\lambda t})}{A e^{\lambda t}(2\lambda+3Ae^{\lambda t})}.
\end{eqnarray}

Similarly, for the hyperbolic model, the statefinder pair are obtained as 

\begin{eqnarray}
r &=& 1+\frac{\mu(\mu+3 \mathcal{H}_0 \sinh{\mu t})}{\mathcal{H}_0^2 \cosh^2{\mu t}},\\
s &=& -\frac{2}{3}\times\frac{\mu(\mu+3\mathcal{H}_0 \sinh {\mu t})}{\mathcal{H}_0(2\mu \sinh {\mu t}+3\mathcal{H}_0 \cosh^2{\mu t})}.
\end{eqnarray}

%*******************************************
\begin{figure}[!h]
\includegraphics[width=105mm]{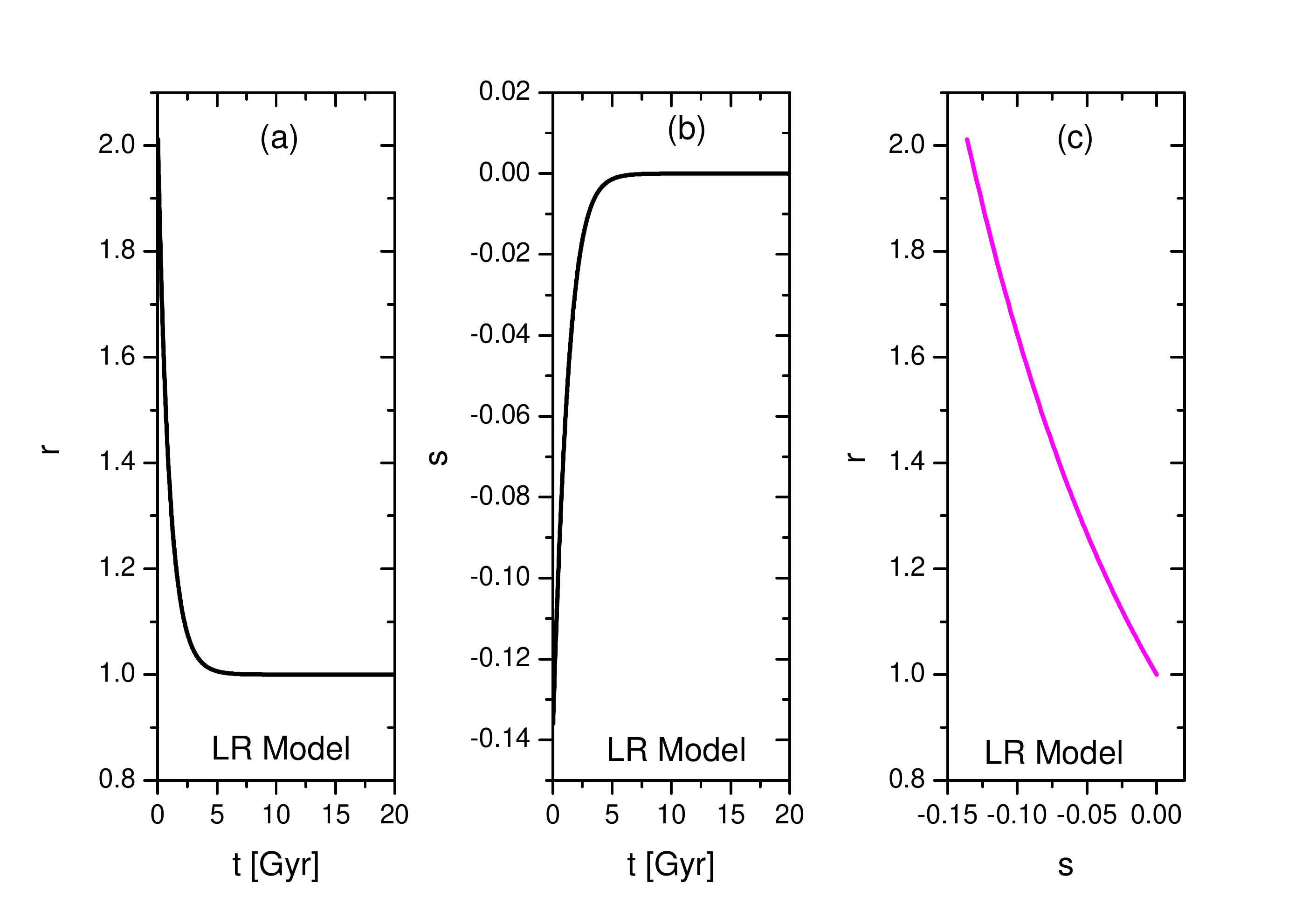}
\caption{The evolution of the state finder pair for the LR model.}
\end{figure}
%**************************************************
\begin{figure}[!h]
\includegraphics[width=105mm]{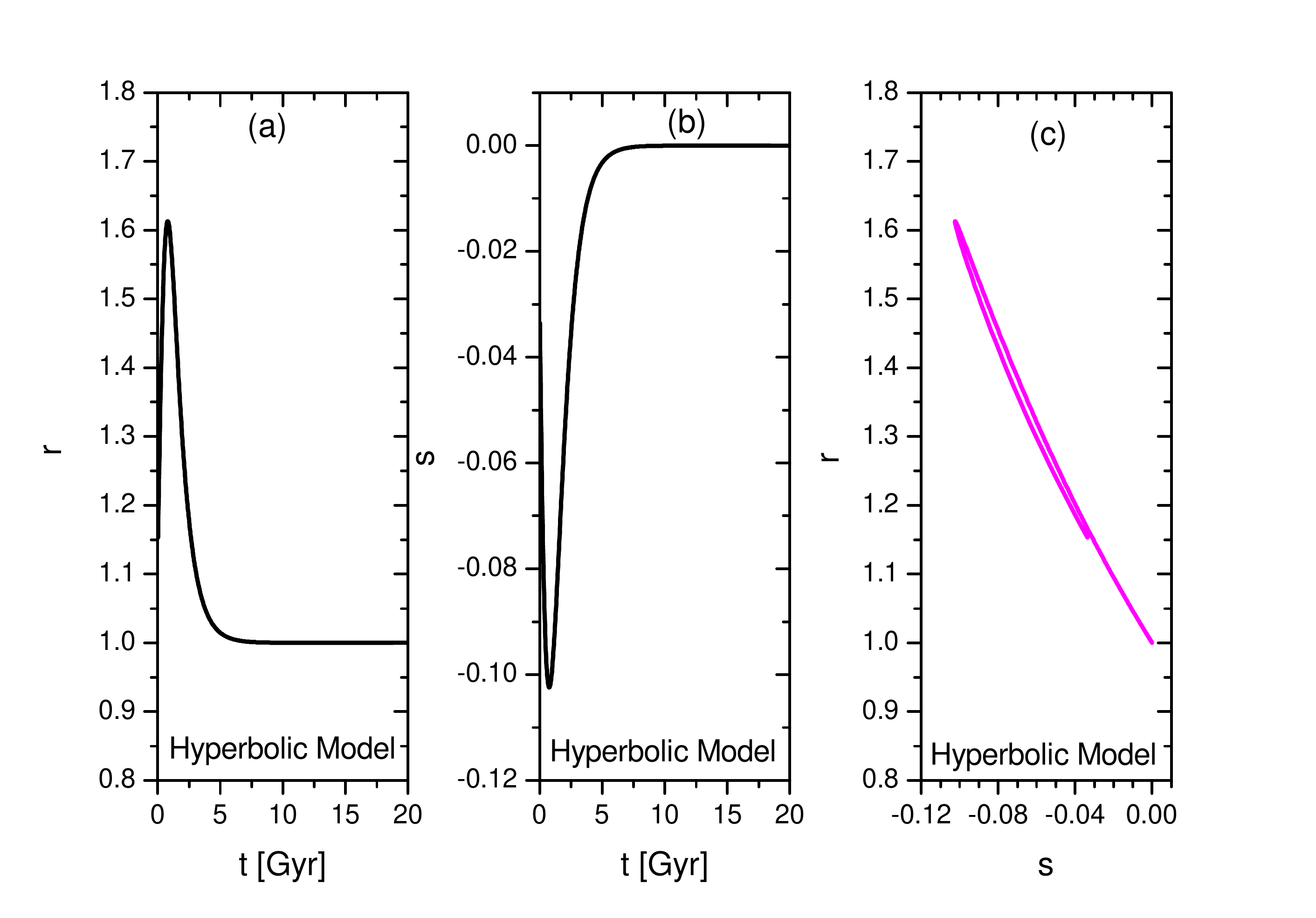}
\caption{The evolution of the state finder pair for the hyperbolic model.}
\end{figure}

For both the models, the statefinder pair evolve with cosmic time. In FIG. 9 and FIG. 10, the evolution of the statefinder pair are shown. Also, we have shown the evolution trajectories of the models in the $r-s$ plane. In case of the LR model (FIG.9), the jerk parameter $r$ being a positive quantity  decreases to become $r=1$ at late times. On the otherhand, the snap parameter $s$ being a negative quantity increases to become $r=0$ at late times. For $\Lambda$CDM model, the statefinder pair has the value $(1,0)$. In otherwords, the statefinder pair $(r,s)$ overlaps with that of the  $\Lambda$CDM model at late time of cosmic evolution. The $r-s$ trajectory shows the evolution of the model towards the $\Lambda$CDM model and thereby confirms its viability. At the present epoch, we found that the present LR model provides $(1, 0)$. 

In the hyperbolic model (FIG.10), the jerk parameter $r$ is a positive quantity. Initially it increases with cosmic time and after attaining a maximum again it decreases to its asymptotic value of $r=1$ at late cosmic phase. The behaviour of the snap parameter $s$ appears to be just opposite to that of the $r$ parameter. $s$ being a negative quantity initially decreases and after attaining a minimum again rises to its asymptotic value $s=0$. The value of the statefinder pair as predicted by the hyperbolic model is $(1, 0)$. The $r-s$ trajectory in FIG.10(c) shows that, this model also evolves to overlap with the $\Lambda$CDM model at late phase of cosmic evolution.

\section{Conclusion}
In this paper, we have presented two cosmological models of the Universe in the framework of $f(R,T)$ gravity theory, where we have considered a minimal coupling between the matter and geometry appearing inside the gravitational action. We chose an anisotropic $BVI_{-1}$ metric for our investigation.  Two different models one with little rip behaviour and the other with a hyperbolic form of the Hubble parameter are constructed. The models provide accelerating behaviour of the Universe at late time of the evolution. For the both the model, we have discussed the dynamical behaviour of the EoS parameter.  More or less, the physical behaviour of both models appear to be the same at least at late times.  However, at an initial phase, the hyperbolic model shows some interesting behaviour. It is observed that, both the models evolve in the phantom-like cosmic phase and at late times overlap with $\Lambda$CDM model. The coupling constant $\gamma$ of the $f(R,T)$ gravity theory affects the dynamics of the models in deciding the path history of the EoS parameter. The effect of $\gamma$ is displayed on the behaviour of the EoS parameter in the sense that less is the value of $\gamma$, higher is the stiffness. However at late phase, all the possible values of $\gamma$ considered in the present work, provide almost similar results for the EoS parameter.  From an analysis of the energy conditions, we found that, the models violate the Strong energy condition, Null energy condition and Weak energy condition and satisfy only the Dominant energy condition. Such violation confirms the phantom-like behaviour of the models. We have carried out a geometric diagnostic test for the LR and hyperbolic model which further confirms the viability of the models that evolve to behave like $\Lambda$CDM model at late phase. As a final remark, we say that, the interesting behaviour of the hyperbolic model at least in the initial phase of cosmic evolution need further investigation concerning the bouncing behaviour or avoiding the Big Bang singularity problem or in getting a transitioning Universe from a decelerated phase of expansion to an accelerated one.

\section*{Acknowledgement}
BM and SKT thank IUCAA, Pune (India) for hospitality and support during an academic visit where a part of this work is accomplished. BM acknowledges DST, New Delhi, India for providing facilities through DST-FIST lab, Department of Mathematics, where a part of this work was done. The authors are thankful to the anonymous referees for their valuable suggestions and comments for the significant improvement of the paper.

\end{document}